\documentclass[pra,twocolumn,superscriptaddress,showpacs,floatfix]{revtex4-1}
\usepackage{graphicx}
\usepackage{dcolumn}
\usepackage{bm}
\usepackage{amsmath,amsfonts,amssymb}

\newcommand{\ket}[1]{\left| #1 \right>} 
\newcommand{\bra}[1]{\left< #1 \right|} 
\renewcommand{\v}[1]{\ensuremath{\mathbf{#1}}} 
\DeclareMathOperator{\sech}{sech}
\DeclareMathOperator{\cotan}{cotan}

\begin{document}
\title{Three-level superadiabatic quantum driving}
\author{Luigi Giannelli}
\affiliation{Dipartimento di Fisica, Universit\`a di Pisa, Largo Pontecorvo 3, 56127 Pisa, Italy}
\author{Ennio Arimondo}
\affiliation{Dipartimento di Fisica, Universit\`a di Pisa, Largo Pontecorvo 3, 56127 Pisa, Italy}
\affiliation{INO-CNR,  Universit\`a di Pisa, Largo Pontecorvo 3, 56127 Pisa, Italy}

\date{February 1, 2014}

\begin{abstract}
  The superadiabatic quantum driving, producing a perfect adiabatic transfer
  on a given Hamitonian by introducing an additional Hamiltonian, is
  theoretically analysed for transfers within a three-level system. Our
  starting point is the stimulated Raman adiabatic passage, realized through
  different schemes of laser pulses. We determine the superadiabatic
  correction for each scheme. The fidelity, robustness and transfer time of
  all the superadiabatic transfer schemes are discussed. We derive that all
  superadiabatic corrections are based on a $\pi$- (or near-$\pi$)-area pulse
  coupling between the initial and final states. The benefits in the protocol
  robustness overcome the difficulties associated to the actual implementation
  of the three-level superadiabatic transfer.
\end{abstract}


\maketitle

\section{Introduction}
The ability to accurately control a quantum system is a fundamental requirement in many areas of modern science  ranging from quantum information processing and coherent
manipulation of quantum systems, to high precision measurements. Very often the quantum control aims at reaching a given target state, as in the preparation of a given atomic/molecular state, or the cooling of atomic ensembles and nano-mechanical
oscillators. The optimum strategy designed for a given task complies with the requests of a nearly perfect fidelity of the final state, of a operation speed  close to the  quantum speed natural lower bound limit rooted in the Heisenberg uncertainty principle, and finally of robustness against imperfections in the quantum control protocol.\\
\indent Superadiabatic~\cite{Lim1991,Berry2009} or transitionless~\cite{Demirplak2003,DemirplakRice2005,Demirplak2008} protocols  and shortcuts to adiabaticity~\cite{Torrontegui2013} have recently received a large attention for the realization of the above targets. In the  superadiabatic/transitionless protocols the controlled system follows perfectly  the instantaneous adiabatic ground state of a given Hamiltonian following the application of an ad-hoc additional Hamiltonian. The shortcuts to adiabaticity are based on the preparation of the controlled system into eigenstates of the Hamiltonian invariants, that characterise all the transformations of the given Hamitonian. The  superadiabatic protocols were recently tested on two-level systems, an energy level  anticrossing for a Bose-Einstein condensate loaded into an optical lattice~\cite{Bason2012,Malossi2013a,Malossi2013b} and the magnetic resonance of a one-half spin~\cite{ZhangSuter2013}. Those experiments demonstrated that superadiabatic protocols realize quantum fidelity equal to one, speed close to the quantum limit, and robustness against parameter variations, making them useful for practical applications. Protocols based on the Hamiltonian invariants have not been tested experimentally so far. \\
\indent The theoretical treatment for the superadiabatic transformations~\cite{Berry2009} provides a quite general approach valid for any multilevel system. The present work applies the superadiabatic transformation to the population transfer in a three-level system. The process of stimulated Raman adiabatic passage (STIRAP) allows to produce an adiabatic passage with the use of two-constant frequency suitably delayed laser pulses~\cite{Bergmann1998,KralShapiro2007}. The high fidelity request is achieved using large pulse areas, i.e., large average Rabi frequencies and long interaction times. For several applications this requirement is a critical disadvantage.  Superadiabatic STIRAP (sa-STIRAP) protocols allow a perfect three-level transfer without the need of intense pulses or long transfer times.\\ 
\indent We determine  several sa-STIRAP protocols, generalising the previous investigations of Demirplak and Rice~\cite{Demirplak2003,DemirplakRice2005} and of Chen {\it et al.}~\cite{ChenMuga2010}. The fidelity of the three-level STIRAP and sa-STIRAP protocols, their robustness against variations of the different parameters, their transfer speeds are calculated. Our analysis demonstrates  that any sa-STIRAP configuration requires the  application of an additional pulse having a $\pi$ (or near-$\pi$) area producing a direct transfer between  the initial and final states, as stated in refs.~\cite{Unanyan1997,ChenMuga2010,TorosovVitanov2013} for Gaussian laser pulses. We demonstrate that a very high fidelity can be reached even releasing that requirement. The analysis of fidelity and robustness is applied also to another scheme of three-level transfer based on the Hamiltonian invariants recently developed by Cheng and Muga~\cite{ChenMuga2012}. Our attention is focused on quantum computation applications of three-level systems where fidelities close to one within one part in a thousand is required. \\
\indent Sect. II introduces the three-level system, the standard STIRAP laser scheme, and also the protocol developed in ref.~\cite{ChenMuga2012} based on the Hamitonian invariants.   Sect. III derives the sa-STIRAP Hamiltonian.  The realisation of the Hamiltonian matrix element connecting the initial and final states required to implement the sa-STIRAP is also presented there. Sect. IV reports numerical analyses of the important features, fidelity, losses, robustness and transfer-time for all the previous introduced  protocols. Sec. V concludes our work.

\section{Hamiltonian and STIRAP}
\subsection{Three level Hamiltonian}
The basic STIRAP process involves three quantum states, linked by two
time-dependent interactions to be referred to as pump, at frequency $\omega_p$
between state $\ket{1}$ of energy $E_1$ and the excited state $\ket{2}$ having
energy $E_2$, and Stokes interaction, at frequency $\omega_s$, between the
intermediate state $\ket{2}$ and the final target state $\ket{3}$ having
energy $E_3$. Two different energy level configurations will be examined, the
ladder one with $E_3>E_2$ and the $\Lambda$ one with $E_3<E_2$.  The
Hamiltonian within the rotating wave approximation
\cite{shore1990t,Bergmann1998}) reads as in the following:
\begin{equation} \label{eq:stirapmatrix} H^0(t)=\frac{\hbar}{2}
  \begin{pmatrix}
    0 & \Omega_p(t) & 0 \\
    \Omega_p(t) & 2 \Delta_p & \Omega_s(t) \\
    0 & \Omega_s(t) & 2 \Delta_3
  \end{pmatrix},
\end{equation}
with $\Omega_p$ and $\Omega_s$ the pump and Stokes Rabi frequencies.  The
detunings from resonance are defined by $\Delta_p = \omega_p -
(E_2-E_1)/\hbar$, $ \Delta_s = \omega_s - (|E_3-E_2|)/\hbar$, and
$\Delta_3 =   \Delta_p + \Delta_s$ for the ladder configuration, and $\Delta_3 =   \Delta_p - \Delta_s$ for the $\Lambda$ configuration.\\
\indent Since the phases of the pump and Stokes fields can be included into
the $\ket{1}$ and $\ket{3}$ definitions without loss of generality, the Rabi
frequency will be chosen real, except in Appendix A. We define an effective
Rabi frequency $\Omega_0$, denoted as adiabatic energy
in~\cite{LaineStenholm1996},
\begin{equation}
  \label{eq:usefuldef}
  \Omega_0(t) = \sqrt{\Omega_p(t)^2+\Omega_s(t)^2}.
\end{equation}
Population losses with $\Gamma_2$ rate from the $|2\rangle$ intermediate level will be also introduced in our analysis.\\
\indent For the transfer process of our interest it is essential to apply the
$\Delta_3 = 0$ two-photon resonance condition, and this case will be here
examined.  The three-level analysis can be written in a simpler form by
defining
\begin{equation}\label{eq:usefuldef2}
  \begin{aligned}
    \tan{\theta(t)} &=\frac{\Omega_p(t)}{\Omega_s(t)}, \\
    \tan{\phi(t)}
    &=\frac{\Omega_0(t)}{\Delta_p+\sqrt{\Delta_p^2+\Omega_0(t)^2}}, \\
  \end{aligned}
\end{equation}
with
\begin{equation}
  \dot{\theta}(t) = \frac{\dot{\Omega}_p(t)\Omega_s(t) - \Omega_p(t)\dot{\Omega}_s(t)}{\Omega_0(t)^2}.
  \label{eq:dottheta}
\end{equation}
The Hamiltonian eigenvalues are written as
\begin{equation} \label{eq:stirapeigenvals2}
  \begin{aligned}
    \lambda_0(t) & = 0 \\
    \lambda_-(t) &=
    -\frac{\hbar}{2}\Omega_0(t) \tan{\phi(t)} \\
    \lambda_+(t) & =
    \frac{\hbar}{2}\Omega_0(t) \cot{\phi(t)} \\
  \end{aligned}
\end{equation}
and the instantaneous eigenvectors are~\cite{Bergmann1998,Fewell1997}
\begin{equation} \label{eq:stirapeigenvect2}
  \begin{aligned} 
    |a_0(t)\rangle&=
    \begin{pmatrix}
      \cos{\theta(t)} \\ 0 \\ -\sin{\theta(t)}
    \end{pmatrix}, \\
    |a_-(t)\rangle&=
    \begin{pmatrix}
      \sin{\theta(t)}\cos{\phi(t)} \\ -\sin{\phi(t)} \\
      \cos{\theta(t)}\cos{\phi(t)}
    \end{pmatrix}, \\
    |a_+(t)\rangle&=
    \begin{pmatrix}
      \sin{\theta(t)}\sin{\phi(t)} \\ \cos{\phi(t)} \\
      \cos{\theta(t)}\sin{\phi(t)}
    \end{pmatrix}.
  \end{aligned}
\end{equation}
As the three-level key feature, the $\ket{a_0(t)}$ eigenstate is a dark state
with zero projection on state $\ket{2}$.
\begin{table*}
  \caption{Temporal dependencies of $\Omega_p/\Omega_{peak}$ and $\Omega_s/\Omega_{peak}$ in STIRAP schemes, and of $\Omega$ in sa-STIRAP schemes. In last column the $\epsilon_1,\epsilon_2$ deviations of $\Omega_d$ from a perfect $\pi$ area pulse. For $\sin^4$  pulses $\Omega_p$ and $\Omega_s$ are different from 0 in the $(\tau<t<\tau+T)$ and $(-\tau<t<T-\tau)$ intervals, respectively; for $\sin/\cos$ both for  $(0<t<T)$.}
  \begin{tabular}{cccccc}
    \hline
    \hline
    $\Omega_p(t)/\Omega_{peak}$  & $\Omega_s(t)/\Omega_{peak}$ &$\Omega_d(t)$& $(\epsilon_1,\epsilon_2)$&Ref.  &   \\
    \hline
    $e^{-\left(\frac{t-\tau}{T}\right)^2}$ & $e^{-\left(\frac{t+\tau}{T}\right)^2}$
    &$4\tau\left[T^2\mathrm{cosh}\left(\frac{4\tau t}{T^2}\right)\right]^{-1}$&(0,0)&\cite{LaineStenholm1996}& \\
    $(1+e^{-t/T})^{-\frac{1}{2}}$ & $(1+e^{+t/T})^{-\frac{1}{2}}$ &
    $\left[{2T \cosh{\left(\frac{t}{2T}\right)}}\right]^{-1}$&(0,0)&   \cite{LaineStenholm1996} \\
    $\sin^4\frac{\pi\left(t-\tau\right)}{T}$  & $\sin^4\frac{\pi \left(t+\tau \right)}{T}$ & 
    $\frac{\pi}{T}\sin{\frac{2\tau \pi}{T}}\left( \cos{\frac{2\pi\tau}{T}} - \cos{\frac{2\pi t}{T}} \right)^3
    \left[\sin^8{\left(\frac{\pi}{T}\left(t-\tau\right)\right)}
      +\sin^8{\left(\frac{\pi}{T}\left(t+\tau \right) \right)}\right]^{-1}$&(0,0)&\cite{Fewell1997} &\\
    $\sin{\left[\frac{1}{2}\arctan\frac{t}{T}+\frac{\pi}{4}\right]}$ & 
    $\cos{\left[\frac{1}{2}\arctan\frac{t}{T}+\frac{\pi}{4}\right]}$ &
    $T\left[t^2 + T^2\right]^{-1}$ &(0,0) & \cite{Elk1995,Fleischhauer1999}\\
    $\sin{\left(\frac{\pi t}{2T}\right)}$  & $\cos{\left(\frac{\pi t}{2T}\right)}$  &  $\pi T^{-1}$&(0,0)&\cite{ChenMuga2012,LaineStenholm1996} \\
    \hline
    $\frac{1}{T}\sech\frac{t}{T}$ & $\frac{\alpha}{T}\sqrt{\left(1-\tanh{\frac{t}{T}}\right)}$  & 
    $4\alpha e^{\frac{t}{T}}\left[T\sqrt{2}\left(\alpha^2 +e^{\frac{2t}{T}} \left( 2+\alpha^2 \right)\right) \sqrt{1+e^{\frac{2t}{T}}}\right]^{-1}$&  $(0,\sqrt{2}\alpha)$& \cite{CarrollHioe1990} \\
    $\frac{1}{T}\sqrt{\left(1-\tanh{\frac{t}{T}}\right) }\sech\frac{t}{T}$ &
    $\frac{\alpha}{T}\left(1-\tanh{\frac{t}{T}}\right)$ & 
    $4\alpha e^{\frac{t}{T}}\left[T\sqrt{2}\left(\alpha^2 +e^{\frac{2t}{T}} \left( 2+\alpha^2 \right)\right) \sqrt{1+e^{\frac{2t}{T}}}\right]^{-1}$&  $(0,\sqrt{2}\alpha)$& \cite{CarrollHioe1990} \\
    $\sech{\frac{t-\tau}{T}}$ & $\sech{\frac{t+\tau}{T}}$ &
    $2\sinh \frac{2\tau}{T}T^{-1}\left[ 1+\cosh\frac{2t}{T}\cosh\frac{2\tau}{T}\right]^{-1}$ &$(e^{\frac{-2\tau}{T}}e^{\frac{-2\tau}{T}})$& \cite{LaineStenholm1996}  \\
    \hline
    \hline
  \end{tabular}
\end{table*}
\begin{figure}[!ht]
  \centering
  \includegraphics[width=\columnwidth]{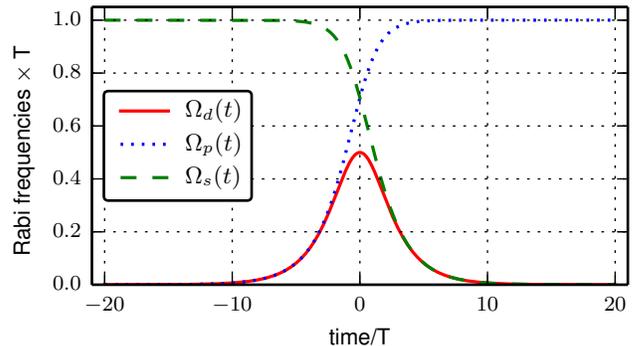}
  \caption{(Color online) Time dependencies of the $\Omega_p$ and $\Omega_s$
    STIRAP exponential pulses (line 2 of Table I) and of the $\Omega_d$
    sa-STIRAP correction. $\Omega_{T}=1$ in the dimensionless units of
    Eq.~\eqref{eq:dimensionless}.}
  \label{fig:1}
\end{figure}

\subsection{STIRAP}
The STIRAP protocol allows to produce an efficient transfer from $\ket{1}$ as initial state to $\ket{3}$ as final state~\cite{Bergmann1998,KralShapiro2007} following the evolution of the $a_0$ dark state. Within the $(t_i,t_f)$ time interval the pump/Stokes lasers are applied as pulses satisfying the well known counterintuitive sequence with $\Omega_s$ first and $\Omega_p$ later, as in Fig.~\ref{fig:1}.\\
\indent We consider the  STIRAP pulses listed in top part of Table I, most of them written as 
\begin{equation}
  \label{eq:typstiappulse}
  \begin{aligned}
    &\Omega_p(t) = \Omega_{peak}f(\frac{t-\tau}{T}) \\
    &\Omega_s(t) = \alpha \Omega_{peak}f(\frac{t+\tau}{T}) \\
  \end{aligned}
\end{equation}
where $f(t)$ is a pulse envelope having unit maximum value, $\Omega_{peak}$ the
peak Rabi Frequency, $2\tau$ the delay between pulses, and $T$ the pulse width. $\alpha$ is a scaling parameter, smaller than 1, introduced for two protocols of Table I where the $\Omega_s$ maximum  is smaller than the $\Omega_p$ one.  The counterintuitive sequence condition imposes $\tau>0$.  Table I includes the $\sin/\cos$ protocol introduced in ref.~\cite{LaineStenholm1996} and rederived in ref.~\cite{ChenMuga2012} through the Hamiltonian invariant approach. Few protocols, for instance the exponential of line 2 and the $\sin/\cos$ of line 4, have finite Rabi frequencies applied at initial or final times. For these protocols, outside the STIRAP time interval   $\Omega_s$ and $\Omega_p$ should be adiabatically switched on/off, respectively. \\
\indent The Rabi frequencies of the above protocols satisfy the following relations 
\begin{equation}
  \begin{aligned}
    \label{eq:stirapcond}
    \lim_{t \to t_i}\frac{\Omega_p(t)}{\Omega_s(t)}&=\lim_{t \to t_i}\tan\theta(t)= \epsilon_1,\\
    \lim_{t \to t_f}\frac{\Omega_s(t)}{\Omega_p(t)}&=\lim_{t \to t_f}\cotan\theta(t) = \epsilon_2,
  \end{aligned}
\end{equation}
with $\epsilon_1,\epsilon_2$ small quantities, equal to zero for protocols on top of the Table I  and different from zero for those on the bottom. The Rabi frequencies temporal dependence implies 
\begin{equation}
  \label{eq:stirapcond3}
  \lim_{t \to t_i}\theta(t) = \epsilon_1, \quad \lim_{t \to t_f}\theta(t) =
 \pm \frac{\pi}{2}-\epsilon_2.
\end{equation}
Thus within the $(t_i,t_f)$ time interval, $\theta(t)$ varies from $\epsilon_1$ to
$\pm\frac{\pi}{2}-\epsilon_2$. For $\epsilon_1=\epsilon_2=0$ the $|a_0(t)\rangle$ dark state varies 
from $|1\rangle$ to $|3\rangle$, while for $\epsilon_1,\epsilon_2 \ne 0$ the $|a_0(t)\rangle$ wavefunction 
contains at initial and final times both $|1\rangle$ and $|3\rangle$ contributions, therefore a small coherence between those states. We have not included in our analysis other protocols based on the presence of a large initial three-level coherence as in~\cite{Fleischhauer1999,ChenMuga2012}, and of higher-order trapping states as in  ref.~\cite{Fleischhauer1999}.\\
\indent  The local adiabaticity condition for a transfer via the $|a_0\rangle$ eigenstate is~\cite{Bergmann1998,FleischhauerManka1996} 
  \begin{equation}
    \label{eq:adiabcond}
    |\dot{\theta}(t)| \ll \frac{1}{2}\left|\Delta_p \pm \sqrt{\Delta_p^2 + \Omega_0(t)^2 }\right|. \\
  \end{equation} 
\indent  By assuming $\Delta_p(t) \ll \Omega_p(t), \Omega_s(t)$, a global adiabaticity condition is derived 
 by time averaging Eq.~\eqref{eq:adiabcond} over the  $\tau$ characteristic time of   the $\Omega_p(t)$ and $\Omega_p(t)$ overlap. For the pulses of Eq.~\eqref{eq:typstiappulse} using Eq.~\eqref{eq:stirapcond3}  the global condition becomes
  \begin{equation}
    \label{eq:adiabglob3}
    \Omega_{peak} \tau \gg 1.
  \end{equation}

\section{sa-STIRAP protocols}
\subsection{Hamiltonian} 
Following refs.~\cite{Lim1991,Berry2009,Demirplak2003,DemirplakRice2005,Demirplak2008} the superadiabatic approach requires the application of a total
Hamiltonian 
\begin{equation}
\label{eq:superadiabatic}
H(t) = H^0(t) + H^1(t),
\end{equation} 
 with the super-adiabatic correction  
\begin{equation}
H^1(t)=i\hbar\sum_n{\left[\ket{\partial_tn}\bra{n}-\langle{n}|\partial_tn\rangle\ket{n}\bra{n}\right]}
\label{eq:supadcorrdef}
\end{equation}
determined from the instantaneous eigenvalues 
$|n(t)\rangle=(|a_0\rangle,|a_-\rangle,|a_+\rangle)$.\\
\indent By applying $H(t)$  the system evolution will follow exactly the instantaneous eigenstate of the
$H^0$ STIRAP Hamiltonian.  If the
system initial state is in the dark one, the system
will remain in that dark state at all times. The adiabatic following of the STIRAP $H^0(t)$ eigenstates
 takes place for any choice of the protocol parameters, even 
with very small values of the applied pump and Stoke fields, and in arbitrary short time.\\
\indent Using Eqs.~\eqref{eq:stirapeigenvect2}  the Hermitian $H^1$ Hamiltonian becomes
\begin{equation} 
\label{eq:stirapsupadcorr} 
H^1(t)=\hbar
  \begin{pmatrix}
    0 & i\dot{\phi}(t)\sin{\theta(t)} & i \dot{\theta}(t) \\
    -i \dot{\phi}(t)\sin{\theta(t)} & 0 & -i \dot{\phi}(t)\cos{\theta(t)} \\
    -i \dot{\theta}(t) & i \dot{\phi}(t)\cos{\theta(t)} & 0
  \end{pmatrix}.
\end{equation}
with the matrix elements given by 
\begin{align} \label{eq:stirapsupadcorrexplicit_12}
  \dot{\phi}(t)\sin{\theta(t)}&=\frac{\Omega_p\left(
      \dot{\Delta}_p\Omega_0-\Delta_p\dot{\Omega}_0\right)}{2 \Omega_0
    \left(\Delta_p^2+\Omega_0^2\right)},\\
  \dot{\phi}(t)\cos{\theta(t)} &= \frac{\Omega_s\left(
      \dot{\Delta}_p\Omega_0-\Delta_p\dot{\Omega}_0\right)}{2 \Omega_0
    \left(\Delta_p^2+\Omega_0^2\right)},
\end{align}
and $\dot{\theta}(t)$ given by Eq.~\eqref{eq:dottheta}.  For
$\dot{\Delta}_p(t)=0$ the above equations reduce to those reported in
ref.~\cite{ChenMuga2010}. For real phases of the pump/Stokes Rabi frequencies, all the $H^1$ elements   are purely imaginary;  for  not real phases see Appendix A. \\
\indent The $H^1_{12}(t)$ and $H^1_{23}(t)$ matrix elements represent a
correction to the pump/Stokes pulses. They impose a phase relation between
their Rabi frequencies and modify their temporal dependence. These matrix
elements vanish in the trivial case $\Omega_p(t) = \Omega_s(t) = \Omega_0(t) =
0, \, \Delta_p(t)\neq 0$, or in the more interesting case of
\begin{equation}
  \label{eq:dotphizero}
     \Delta_p(t) =C\cdot \Omega_0(t) 
\end{equation}
Here $C$ is any constant, also zero, leading to the convenient choice of $\Delta_p(t)$ a constant equal to
zero for annulling those corrections.\\
\indent The most interesting $H^1_{13}(t)$ matrix
element  couples
directly the initial and final states. It will be written as 
\begin{equation} 
\label{eq:supadham13_recall} 
H^1_{13}(t) =
  \hbar\frac{i\Omega_d(t)}{2}.
\end{equation}
having defined the \textit{detuning pulse} as
\begin{equation} \label{eq:omegad}
 \Omega_d(t) \equiv
  2\dot{\theta}(t).
\end{equation}
Such definition was introduced in ref.~\cite{Unanyan1997} and analysed in ref.~\cite{Fleischhauer1999} because in the $\ket{a_n(t)}$ basis, 
$\Omega_d(t)$  represents a detuning energy of the $|a_0(t)\rangle$ dark state.  In the $|1,2,3\rangle$ basis $i\Omega_d(t)$ represents a Rabi frequency connecting states $|1\rangle$ and $|3\rangle.$ Because the $H^1$ Hamiltonian is written within the rotating-wave approximation, the $i\Omega_d$ Rabi coupling is created by a field oscillating at frequency $\omega_p-\omega_s$ in the $\Lambda$ configuration, and frequency $\omega_p+\omega_s$ in the ladder one. The $\Omega_d(t)$ functions associated to the different STIRAP pulses are reported in the last column of Table I, and that for the exponential pulses is plotted in Fig. \ref{fig:1}.  \\
\indent If a proper time dependence of $\Omega_p(t)$ and $\Omega_s(t)$ could be
found such that the detuning pulse, i.e., $H^1_{13}(t)$, is
identically zero, a
super-adiabatic evolution of the system could be produced only by changing the
shape of the pump and Stokes fields. But this is
not the case because $\Omega_d(t) = 0$  only if  
\begin{equation} 
\label{eq:supadham13nonzero} 
\Omega_p(t) \propto \Omega_s(t).
\end{equation}
For this non-counterintuitive configuration    the $| a_0(t)\rangle$
dark state  does not
link anymore the $| 1\rangle$ and $| 3\rangle$ states. Appendix A deriving $\Omega_d(t)$ in presence of a time dependence of the pump/Stokes field phases  demonstrates that also in that case the $H^1_{13}(t)= 0$ request cannot be satisfied. While the key role of $\dot{\theta}$ in controlling the nonadiabaticity of STIRAP was pointed out by several authors, see~\cite{LaineStenholm1996,Fleischhauer1999},  we derive that the $\dot{\theta}$ Rabi frequency coupling between initial and final states is strictly required in the sa-STIRAP realisation.\\
\indent The time dependence of $\Omega_d(t)$, {\it i.e.}, $\dot{\theta}(t)$, is determined  
by Eq.~\eqref{eq:dottheta}. Using Eq.~\eqref{eq:stirapcond3} for our STIRAP pulses, that time dependence leads to 
\begin{equation}
  \label{eq:omegadpipulse}
  \int_{t_i}^{t_f}\Omega_d(t)dt = 
  \int_{t_i}^{t_f}2\dot{\theta}(t)dt 
  =\pm \pi- 2(\epsilon_1+\epsilon_2).
\end{equation}
\indent Therefore the $\Omega_d(t)$ dependence corresponds to a nearly $\pi$-area pulse, a perfect one for several pulses, for instance the Gaussian one, as shown by the $\epsilon_1,\epsilon_2$ values in last column of Table I. Notice that a resonant $\pi$-area pulse connecting the states $|1\rangle$ and $|3\rangle$ produces a
complete population transfer by itself. Thus we obtain a deceiving result: the superadiabatic 
STIRAP realisation implies the application of an additional electromagnetic field that in the $\pi$-area pulse configuration  
produces by itself the required transfer. As investigated in the following, the combination of STIRAP and near $\pi$-area  detuning pulses becomes useful when examining the robustness of the different transfer schemes. The $\pi$-area pulse requirement was derived in~\cite{Unanyan1997,Fleischhauer1999}  while searching for an improvement of the
STIRAP efficiency by adding a low frequency field. We have demonstrated that the near $\pi$-area pulse condition derives from 
the superadiabatic construction and leads to the complete cancellation of the non adiabatic losses.\\ 

\begin{figure}
  \centering
  \includegraphics[width=\columnwidth]{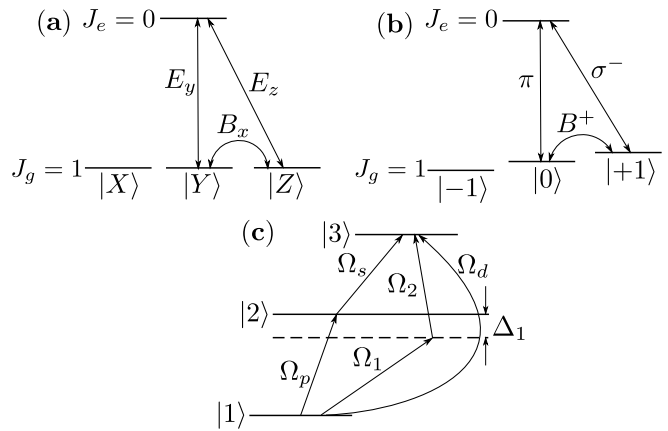}
  \caption{In (a) and (b) sa-STIRAP realisation in a $\Lambda$ system between
    the $J_g=1$ Zeeman sub-levels and an excited $J_e=0$ state. In (a) laser
    configuration defined within a Cartesian $|i>,\,(i=X,Y,Z)$ basis set with
    linearly polarized lasers and static magnetic field $B_x$. In (b) laser
    configuration defined in the basis of the $J_z$ eigenstates with $\pi$/
    $\sigma^-$ lasers and a $\sigma^+$ circularly polarized magnetic
    radiofrequency field.  In (c) sa-STIRAP in a ladder system with the the
    detuning pulse produced by an additional two-photon transition.}
  \label{fig:LambdaLadder}
\end{figure}

\subsection{Detuning pulse realization}
\subsubsection{Magnetic-dipole in $\Lambda$ scheme} \label{sec:magn-dipole-trans} 
Owing to the parity selection rules, in
the $\Lambda$ systems of Figs.~\ref{fig:LambdaLadder}(a)  and (b) the $i\Omega_d$ coupling
between the $\ket{1}$ and $\ket{3}$ states may be originated by a magnetic
dipole interaction  between the $\v{J}$ atomic/molecular angular momentum and
an external magnetic field $\v{B}$ 
\begin{equation}
  \label{eq:hammagnetic}
  H_B = \mu_B g_J \v{J} \cdot \v{B}
\end{equation}
$\mu_B$ being the Bohr magneton and $g_J$ the Land\'e factor. The detuning pulse is
\begin{equation}
  \label{eq:magneticmatrixelement}
 \frac{i\Omega_d(t)}{2} = \mu_B g_J \frac{\bra{1} \v{J} \cdot \v{B}(t)\ket{3}}{\hbar},
\end{equation}
 \indent As proposed in ref.~\cite{Unanyan1997} that magnetic coupling may be realized  for a $J_g = 1$ ground state 
and  an excited
$J_e = 0$ state.  A $\Lambda$-system  with
two ground Zeeman states and the excited state is selected by properly choosing the laser polarizations.  In  the $\ket{X}$, $\ket{Y}$ and $\ket{Z}$ Cartesian basis of the $J_g=1$ state and for pump/Stokes laser fields linearly polarized along the $y$ and $z$ axis, the scheme works out with $\ket{1}=\ket{Y}$ and 
$\ket{3}=\ket{Z}$, as in Fig.~\ref{fig:LambdaLadder}(a).  A $B_x$ magnetic field along the $x$ axis produces the imaginary detuning pulse
\begin{equation}
  \label{eq:Jximag}
  \bra{Y}{H_{13}}\ket{Z} =i\hbar \frac{\Omega_d(t)}{2}= -i \mu_B g_J B_x
\end{equation}
\indent Another simple realization based on the spherical basis of the $J_g=1$ state is shown in Fig.~\ref{fig:LambdaLadder}(b) in presence of a magnetic splitting between the $\ket {J_g,m_J=0}$ and  $\ket {J_g,m_J=1}$ Zeeman sublevels. In this case the pump/Stokes fields have $\pi$  and $\sigma^-$ polarizations and  the detuning pulse is based on 
a $\sigma^+$ circularly polarized resonant radiofrequency magnetic field $\v{B}$ in the $x-y$ plane. This field phase should be $\pi/2$ shifted in respect to that of $\omega_p-\omega_s$. The generalisation of these schemes to different atomic configurations is not obvious.  
\subsubsection{Two-photon transition in ladder} \label{sec:two-phot-trans} In
a ladder level scheme the direct interaction between states $\ket{1}$ and
$\ket{3}$ may take place via a two-photon transition, as in
Fig.~\ref{fig:LambdaLadder} (c).  Two new lasers with Rabi frequencies
$\Omega_1(t)$ and $ \Omega_2(t)e^{i\frac{\pi}{2}}$ are added to the system in
order to satisfy the two-photon resonance. The one-photon resonance is detuned
by $\Delta_1$ from the $|2\rangle$ intermediate state. We impose the sa-STIRAP
$H^1_{13}(t)$ matrix element of Eq.~\eqref{eq:supadham13_recall} equal to the
two-photon Rabi frequency~\cite{BrionMolmer2007} and obtain
\begin{equation}
  \label{eq:omegadviaomegar} 
  i\frac{\Omega_d(t)}{2} = -\frac{\Omega_1(t) \Omega_2(t)e^{-i\frac{\pi}{2}}}{2\cdot 2\Delta_1}=i\frac{\Omega_1(t) \Omega_2(t)}{4 \Delta_1}.
\end{equation}
At fixed $\Delta_1$ a simple choice is to $\Omega_1(t) =\Omega_2(t)=
\sqrt{2\Delta_1\Omega_d(t)}.$ For an experimental realisation of this scheme
two sidebands of the pump and Stokes frequencies at radiofrequencies
$\omega_{rf}$ and $-\omega_{rf}$, respectively, could be created. In addition
by imposing a ninety degree phase shift of the radiofrequency fields the
imaginary sign of the detuning field can be produced.
\begin{figure*}[!ht]
  \centering
  \includegraphics[width=\textwidth]{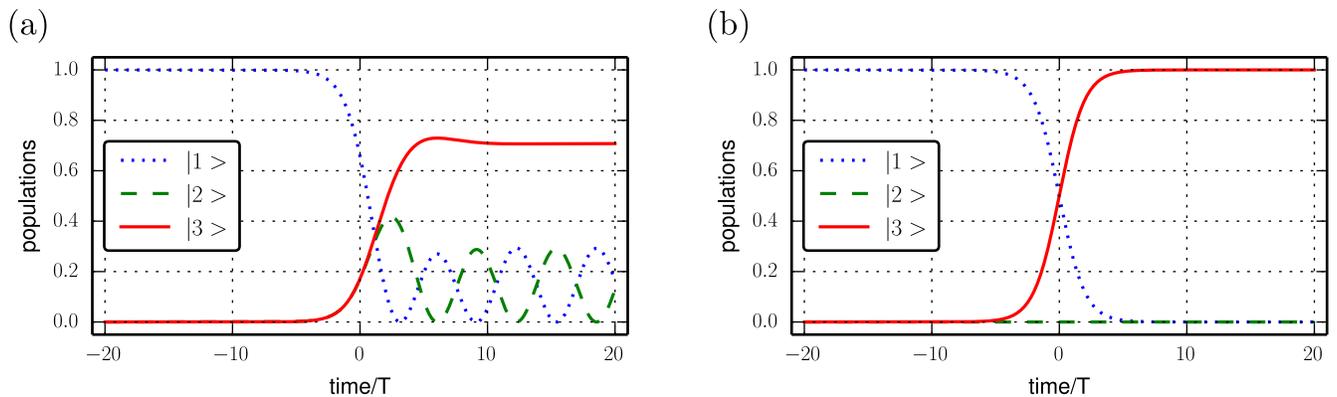}
  \caption{(Color online) Time dependence of the $|1,2,3\rangle$ populations
    for the STIRAP and sa-STIRAP exponential protocols in (a) and (b),
    respectively, for $\Omega_T=1$.  $\Gamma_2=0$ in all cases.}
  \label{fig:exponential}
\end{figure*}

\section{Numerical analyses}
This Section derives  fidelity, robustness, population loss from the intermediate state, robustness and speed-limit for the STIRAP/sa-STIRAP protocols having $\epsilon_1=\epsilon_2=0$.   Four parameters describe the laser  and atom evolutions: i)  the $\Omega_{peak}$peak Rabi frequency,  ii) the $T$ laser pulses duration, iii) the  $2\tau$ delay between pulses  measuring their overlap, and iv) the $\Gamma_2$ decay rate of the $|2\rangle$ state. The system evolution
  is characterised by a  timescale invariance. Then for all values of the previous parameters, except for a timescale factor, 
  the system evolution   is fully defined by three quantities 
  \begin{equation}
\Omega_T= \Omega_{peak}T,\quad \tau_T= \frac{\tau}{T}, \quad \Gamma_T= \Gamma_2 T.
 \label{eq:dimensionless}
  \end{equation}
  That is easily verified for the Gaussian pulses, and it was numerically verified also for other pulses. In the following we will use $T=1$ $\mu$s as reference timescale, and the  presented plots should be properly scaled for analysing other pulse durations. 
\subsection{Fidelity}
The following fidelity characterises completely the $\ket{\Psi(t)}$ wavefunction transfer  to the $\ket{3}$ state:
\begin{equation}
F= |\langle 3| \Psi(t_{fin})\rangle |^2.
\end{equation}
For the $\Gamma_2=0$ case  $F>0.95$ values  are achieved applying a STIRAP  protocol with area parameter  $A=\Omega_{peak}\tau$ around 10-20. Therefore the sa-STIRAP protocol is useful mainly at low values of $A$, where the STIRAP fidelity greatly depends on the $f(t)$ temporal dependence. For the exponential pulses of Fig.~\ref{fig:1} with $\tau_T=\Omega_{T}=1$ STIRAP produces a maximum final fidelity around 0.7, as in Fig.~\ref{fig:exponential}(a). Instead the sa-STIRAP final fidelity is  1 for all the $\tau$, $\Omega_{peak}$ values, as in Fig.~\ref{fig:exponential}(b).\\
\begin{figure}[!ht]
  \centering
  \includegraphics[width=\columnwidth]{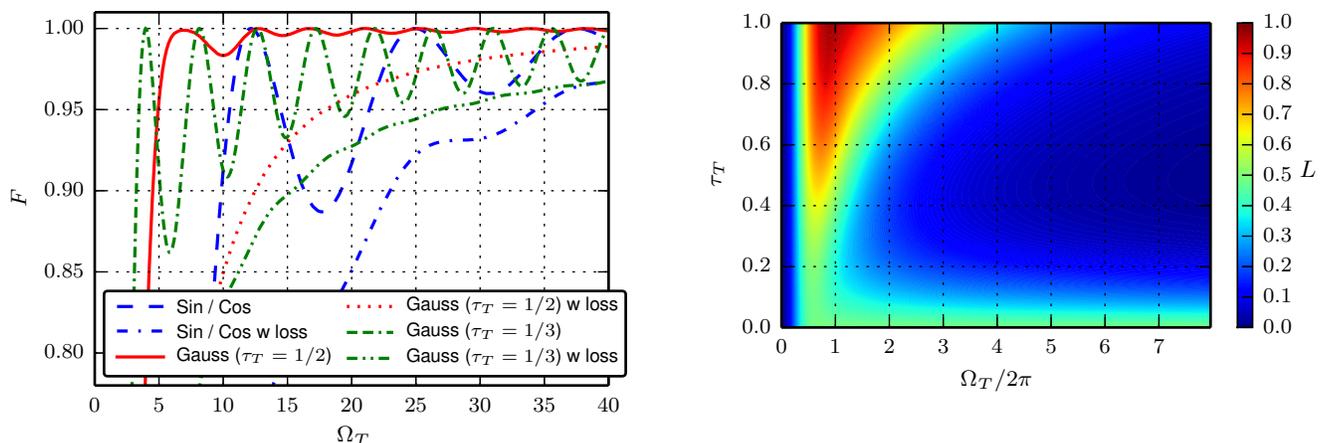}
  \caption{(Color online) Fidelity vs $\Omega_{T}$ for STIRAP Gaussian and
    $\sin/\cos$ protocols of Table I, the Gaussian ones with different values
    of $\tau_T$. No loss ($\Gamma_2=0$) and with loss at $\Gamma_T=4$. }
  \label{fig:chenmuga}
\end{figure}
\indent For a quantum computation target, where  $F>0.999$ fidelities are required, the STIRAP protocol is not good enough, as it appears in Fig.~\ref{fig:chenmuga} for the Gaussian protocol.  Its fidelity is characterised by an oscillating dependence on the $\Omega_T$ parameter, with the required very high fidelities reached only in narrow parameter regions. A similar oscillating dependence occurs also for exponential STIRAP pulses, with reduced oscillation amplitude. For the $\sin/\cos$ transfer protocol  of Table I   Chen et al.~\cite{ChenMuga2012} presented the following oscillating dependence of $F$ on the $\Omega_{peak}$ value:
\begin{equation}
\sqrt{F}=1-\sin^2\epsilon\left[1-\cos\left(\frac{\pi}{1\sin\epsilon}\right)\right],
\end{equation}
with $\epsilon={\mathrm {arccot}}(\Omega_T/\pi)$. This dependence is also shown in Fig.~\ref{fig:chenmuga}. In all cases the $F=1$ maxima appear when the Rabi oscillations between the system levels are in phase with the interaction time. The successive maxima are obtained for an increasing number of Rabi oscillations. A similar oscillating dependence of the fidelity on the protocol parameters was also found in ref.~\cite{Malossi2013a} for a nonlinear Landau-Zener sweep.\\
\begin{figure}[!ht]
  \centering
  \includegraphics[width=\columnwidth]{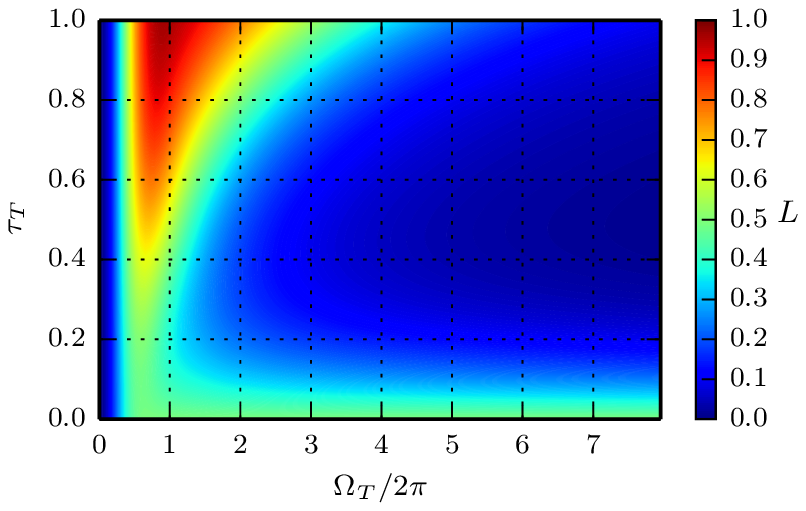}
  \caption{(Color online) $L$ STIRAP losses vs $\tau_T$ and $\Omega_T$ for
    Gaussian pulses with $\Gamma_T=10$.}
  \label{fig:5}
\end{figure}
\indent The introduction of a loss rate for the $\ket{2}$ intermediate state decreases the fidelity   because of losses from that state all along the  temporal evolution, see curves in Fig.~\ref{fig:chenmuga}. \\

\begin{figure*}[!ht]
  \centering
  \includegraphics[width=\textwidth]{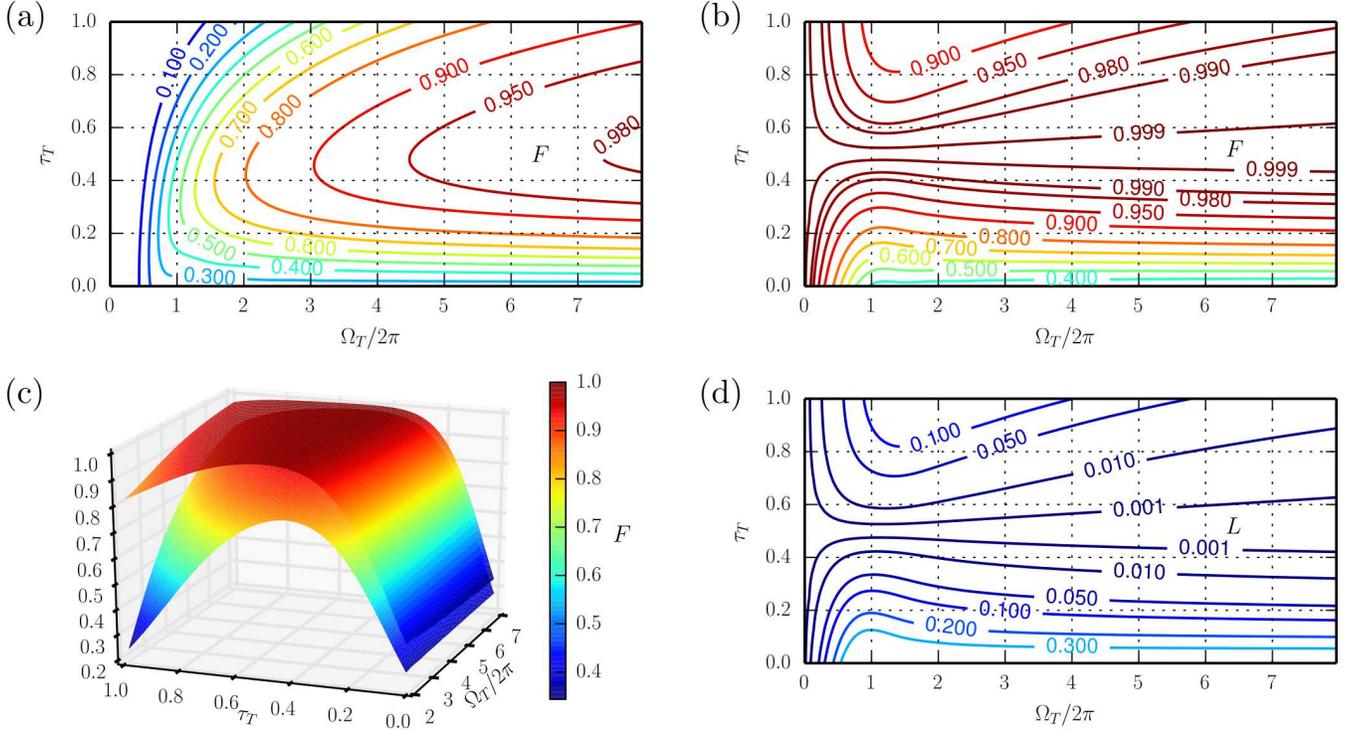}
  \caption{(Color online) $F$ and $L$ vs $\tau_T$ and $\Omega_T$ for STIRAP
    and sa-STIRAP Gaussian pulses at $\Gamma_T=10$. Fidelity $F$ in (a) and (b)
    2D plots, and in 3D surfaces of (c). Losses $L$ in 2D plot (d). In (a) and
    bottom surfaces of (c) for STIRAP protocols; in (b), top surface of (c)
    and (d) for sa-STIRAP protocols.  For the sa-STIRAP's $F$ deviates from
    one and $L$ from zero because the detuning pulse is blocked at its value
    for $\tau_T=1/2$.}
  \label{fig:stiraprobustness}
\end{figure*}

\subsection{Population loss}
In presence of the $\Gamma_2$ loss rate the population loss to internal and external  states during the system evolution is an important parameter of the transfer protocol.
That loss is given by 
\begin{equation}
  \label{eq:intpop2def}
 L =   \Gamma_2 \int_{t_i}^{t_f}\left|{\bra{2}{\Psi(t)}}\right|^2dt.
\end{equation}
 $(1-F)$ represents a good approximation to this quantity, except for population left over in the $\ket{1}$ state. Fig.~\ref{fig:5} reports $L$  results for Gaussian STIRAP transfers. Optimum transfer, i.e. $L$ minimum, is produced for large $\Omega_T$ values and for an optimum $\tau_T$.  These losses are totally eliminated in sa-STIRAP protocols. \\

 \subsection{Robustness}
\begin{figure}[!ht]
  \centering
  \includegraphics[width=\columnwidth]{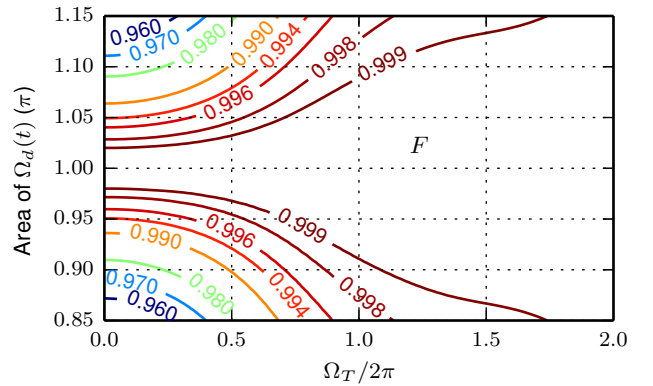}
  \caption{(Color online) 2D plot of the sa-STIRAP fidelity vs $\Omega_{T}$
    and the detuning pulse area around the $\pi$ optimal value, for Gaussian
    pulses with $\tau_T=0.5$ and $\Gamma_T=1$.}
  \label{fig:efficiencypipulsevsaarea}
\end{figure}

In order to test the sensitivity of the STIRAP and sa-STIRAP protocols to a (simulated) variation in the control parameters, we varied the protocol parameters around the optimum value of the ${\cal F}$ fidelity. \\ 
\indent  In  the case of  $\Gamma_2=10/T$, i.e., a $\ket{2}$ decay rate
much  greater than $T^{-1}$, STIRAP and sa-STIRAP results for Gaussian pulses are compared in
Fig.~\ref{fig:stiraprobustness} as function of $\tau_T$ and $\Omega_T$. The STIRAP numerical results  are
 in Fig.~\ref{fig:stiraprobustness}(a) and bottom surface of (c).  The $F>0.999$ requirement is not satisfied in the explored range of delay and  Rabi frequency values. The (a) and (b) plot  evidences  out that in the presence  of the $\Gamma_2$ loss all the fidelity contours shrink in space and a high fidelity requires larger Rabi frequencies.  Similar $F$ tests for the sa-STIRAP protocol, assuming the application of the $\Omega_{d}$ pulse, are in  Fig.~\ref{fig:stiraprobustness}(b) and top surface of (c), while $L$ losses are plotted in (d). In all these plots the detuning pulse remains locked at  its value for $\tau_T=1/2$, and that demonstrates that very high fidelities, and very small losses, are reached even if the detuning pulse is not perfectly matched at the value given by Eqs.~\eqref{eq:omegad} and \eqref{eq:dottheta}. As a test of the sa-STIRAP robustness, from the area in Fig.~\ref{fig:stiraprobustness}(b) we derive that the  the $F>0.999$ condition is verified when
 \begin{equation}
 \Delta\tau/\tau<0.35,\quad \Omega_T>2.
 \end{equation}
 \indent In  Fig.~\ref{fig:efficiencypipulsevsaarea} the fidelity is plotted varying both the detuning pulse area and the pump/Stokes maximum Rabi
frequency. Those results show   that the presence of the pump and Stokes
pulses makes the  $\pi$-area pulse more robust
against  a variation of its area. Finally testing under the same conditions the robustness of the detuning pulse phase, we found that the $F>0.999$ condition is verified for a detuning phase different from $\pi$ less than $\pi/40$ in the worst condition of $\Omega_{T} \approx 0.7$. At smaller and larger $\Omega_{T}$ values the detuning pulse and the pump/Stokes  pulses, respectively, dominate the system evolution and the stability conditions are more relaxed. \\
\indent The plots of Fig. \ref{fig:chenmuga} indicates the limited robustness of the Gaussian and $\sin/\cos$ STIRAP protocols produced by the oscillating behaviour: the $F>0.999$ condition requires a $\Omega_T$ control around three percent.\\

\subsection{Protocol quantum speed}
\begin{figure}[ht]
  \includegraphics[width=\columnwidth]{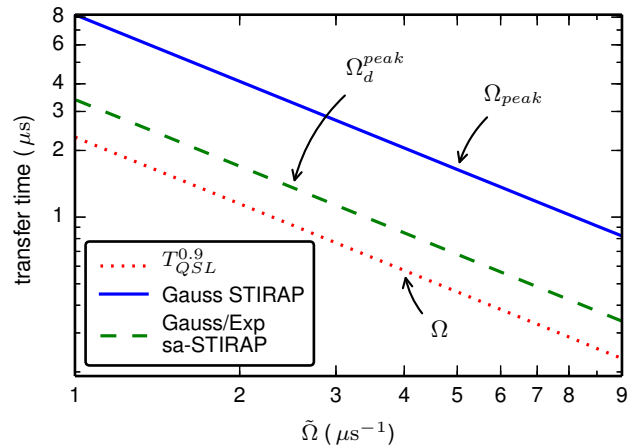}
  \caption{(Color online) Three-level $T^{0.9}$ transfer time vs a
    characteristic $\tilde{\Omega}$ Rabi frequency. In the top vs
    $\tilde{\Omega}=\Omega_{peak}$ for the Gaussian STIRAP protocol with
    optimized $T$ and $\tau$ parameters. In the center line vs
    $\tilde{\Omega}=\Omega_d$ peak value for the Gaussian/exponential
    sa-STIRAP protocols. In bottom line vs the $\tilde{\Omega}=\Omega $ Rabi
    frequency of a $\ket{1}$ and $\ket{3}$ direct coupling for the quantum speed
    limit, this limit reached also by the sa-sin/cos protocol with
    $\tilde{\Omega}=\Omega_d$. $\Gamma_2$ =0 everywhere.}
  \label{fig:timeprocesslog}
\end{figure}

We investigate here the time required for the three-level quantum transfer. We
define the three-level quantum transfer time $T^{0.9}$ as the time interval
between the ninety-nine percent occupation of the initial and the
ninety occupation of the final $\ket{2}$ state. This definition does not match the standard one with an initial total occupation of the $\ket{1}$ state~\cite{Giovannetti2003}, in order to deal with the STIRAP protocols switched on at $t=-\infty$.  Our results of $T^{0.9}$ vs the $\Omega_{peak}$ value  for the Gaussian protocol are plotted in Fig.~\ref{fig:timeprocesslog}.  For each data point both the $T$ and  $\tau_T$ values were optimised.\\
\indent We compare those data to the quantum speed limit required for the
three-level transfer. The previous analyses on the quantum speed limit in
two-level systems~\cite{Giovannetti2003,Bason2012,Malossi2013a} have
demonstrated that the quantum speed limit is reached by a NMR-type
$\pi$-pulse, with constant Rabi frequency. Taking into account our initial and
final conditions on the state occupation for a $\pi$ pulse we obtain the
following relation between the quantum speed limit time $T^{0.9}_{QSL}$ and
the applied Rabi frequency $\Omega$:
\begin{equation}
  \label{eq:qsl}
  T^{0.9}_{QSL}= \frac{2.29}{\Omega},
\end{equation}
If $\Omega$ is non constant, the denominator of the above equation should be replaced by its time average. $T^{0.9}_{QSL}$ is plotted in Fig. \ref{fig:timeprocesslog} vs the   $\Omega$ amplitude of an electromagnetic field producing the direct transfer between initial and final states given by Eq.~\eqref{eq:omegad}. None STIRAP transfer is faster than the quantum speed limit. The difference between the QSL and STIRAP transfer times greatly depends on the temporal superposition of the pump/Stokes pulses.\\
\indent The temporal evolution of all sa-STIRAP protocols is determined by the
$\Omega_d(t)$ detuning pulse. The associated transfer time is derived from
Eq.~\eqref{eq:qsl} by inserting the time average value of $\Omega_d$
\begin{equation}
  T^{0.9}= \pi\frac{t^f_{0.9}-t^i_{0.99} }{\int_{t^i_{0.99}}^{t^f_{0.9}}\Omega_d(t)dt},
\end{equation}
where $t^i_{0.99}$ and $t^f_{0.9}$ are the initial and final times,
respectively, where fidelity reach those values.  Because of the area relation
of Eq.~\eqref{eq:omegadpipulse} the $\Omega_d$ temporal average is very close
to $\pi$, and therefore the transfer time depends on the temporal evolutions
of $\Omega_d$ reported in column 3 of Table I. Both $T$ and $\tau$ parameters
appear in those evolutions, and for each protocol the transfer time has a
specific dependence on the parameters.  Notice that for the sa-sin/cos
protocol, where a constant $\Omega_d=\pi/T$ pulse is applied (see line 5 of
Table I), the speed-limit of Eq.~\eqref{eq:qsl} is reached. In the case of the
sa-$\sin^4$ protocol the transfer time depends on $\tau_T$ reaching
$T^{0.9}=1.10(3)T^{0.9}_{QSL}$ at $\tau_T=1/15$.  For the Gaussian/exponential
pulses $T^{0.9}=1.48(3)T^{0.9}_{QSL}$, independently of $\tau_T$.  Finally for
the sa-sin/cos(arctan) protocol, line 4 of Table I, not containing the $\tau$
parameter, $T^{0.9}=2.73(5)T^{0.9}_{QSL}$.

\section{Conclusions}
The present work determines the corrections to the STIRAP pulses required to produce a superadiabatic transfer with fidelity equal one in a three-level system.  Each STIRAP protocol has a different superadiabatic correction. The superadiabatic Hamiltonian requires the application of the detuning pulse as a direct coupling between the initial and final states. For the $\Lambda$ level scheme a magnetic field, and for the ladder configuration  a two-photon transition  with laser fields created as radiofrequency sidebands from the pump/Stokes laser beams will produce that coupling.  That direct  interaction should be applied in a $\pi$-area, or near $\pi$-area,  pulse configuration, and the application of the detuning pulse alone could produce the desired transfer. However we demonstrate that the combination of STIRAP and detuning pulse has a robustness much larger than each separate transfer. The sa-STIRAP transfer occurs within a temporal window imposed by the applied detuning pulse.  Transfer times close to a three-level quantum speed limit may be reached.\\
\indent Even if the technical effort, and the energy request,  required to realize a sa-STIRAP protocol may be considerable, its implementation is  very useful in quantum driving realisations with heavy requests of efficiency and stability. The application of additional electromagnetic fields is required, for instance, also in the robust composite stimulated Raman adiabatic passage proposed in ref.~\cite{TorosovVitanov2013}. We conclude that any gain in fidelity and in stability requires additional resources in experimental tools and laser power irradiating the sample.\\

\section{Acknowledgments}
This work was supported by the PRIN-2009 Project Quantum Gases beyond Equilibrium of the MIUR-Italy and by the EU Marie Curie ITN COHERENCE. The authors thank R. Mannella and P. Pillet for useful discussions and suggestions, and D. Ciampini for a careful exam of the manuscript.

\appendix
\section {Detuning pulse with phase control of pump/Stokes pulses}

This Appendix examines the sa-STIRAP Hamiltonian by introducing a temporal dependencies for the phases of  the pump and Stokes Rabi frequencies:
\begin{equation}
  \begin{aligned}
    \Omega_p(t) &= e^{i\phi_p(t)}|\Omega_p(t)|, \\
    \Omega_s(t) &= e^{i\phi_s(t)}|\Omega_s(t)|.
  \end{aligned}
\end{equation} Assuming $\Delta_p(t)=0$, as in Eq.~\eqref{eq:dotphizero}, in order  to
have $H^1(t)_{12}$ and $H^1(t)_{23}$ identically zero, the super-adiabatic Hamiltonian of Eq.~\eqref{eq:superadiabatic} for the Ladder system becomes
\begin{equation}
H(t) = \frac{\hbar}{2}
    \begin{pmatrix}
      0 & \Omega_p^*(t) & i\Omega_d^*(t) \\
      \Omega_p(t) & 0 & \Omega_s^*(t) \\
      -i\Omega_d(t) & \Omega_s(t) & 0
    \end{pmatrix},
    \end{equation}
and for the $\Lambda$ one
\begin{equation}
    H(t) = \frac{\hbar}{2}
    \begin{pmatrix}
      0 & \Omega_p^*(t) & i\Omega_d^*(t) \\
      \Omega_p(t) & 0 & \Omega_s(t) \\
      -i\Omega_d(t) & \Omega_s^*(t) & 0
    \end{pmatrix}. 
\end{equation}
In a definition of the detuning pulse more general than in
Eqs.~\eqref{eq:omegad} and ~\eqref{eq:dottheta}, here
\begin{equation}
  \begin{aligned}
    \label{eq:generalomegad}
    \frac{\Omega_d(t)}{2} &= \frac{\dot{\Omega}_p(t)\Omega_s(t)-\Omega_p(t)
      \dot{\Omega}_s(t)}{|\Omega_p(t)|^2+|\Omega_s(t)|^2} &
    \text{for Ladder}, \\
    \frac{\Omega_d(t)}{2} &= \frac{\dot{\Omega}_p(t)\Omega^*_s(t)-\Omega_p(t)
      \dot{\Omega}_s^*(t)}{|\Omega_p(t)|^2+|\Omega_s(t)|^2} & \text{for
      $\Lambda$.}
  \end{aligned}
\end{equation}
\indent For the Ladder system the detuning may be written as
\begin{equation}
  \begin{aligned}
   \frac{\Omega_d(t)}{2} =
   &e^{i(\phi_p+\phi_s)}\left[|\Omega_p(t)|^2+|\Omega_s(t)|^2\right]^{-1} \\
    &\times \left[\dot{|\Omega}_p(t)||\Omega_s(t)|-|\Omega_p(t)|
      \dot{|\Omega}_s(t)|\right. \\
      &\qquad \left. +i(\dot{\phi_s}-\dot{\phi_p})|\Omega_p(t)||\Omega_s(t)|\right].
  \end{aligned}
\end{equation}
In order to reduce the detuning pulse we obtain an additional condition on the pump and Stokes phases,
$\dot{\phi_s}=\pm\dot{\phi_p}$ in the Ladder/$\Lambda$ schemes respectively, or  $\dot{\phi_s}=\dot{\phi_p}=0$. However  also with that phase control  the sa-STIRAP protocol $\Omega_d$  cannot be identically zero.

%

\end{document}